\documentclass[pra,twocolumn,tightenlines,showpacs,nofootinbib]{revtex4}
\usepackage{bm,dcolumn,amsmath,graphicx,amsfonts,amssymb}

\begin{document}
\title{Ion clock and search for the variation of the fine structure
  constant using optical transitions in Nd$^{13+}$ and Sm$^{15+}$.}

\author{V.A. Dzuba$^{1,2}$, A. Derevianko$^2$, and V. V. Flambaum$^1$}
\affiliation{
$^1$School of Physics, University of New South Wales,
Sydney, NSW 2052, Australia}
\affiliation{
$^2$Department of Physics, University of Nevada, Reno, Nevada 89557, USA}

\begin{abstract}
We study ultranarrow $5s_{1/2}$ - $4f_{5/2}$ transitions in Nd$^{13+}$
and Sm$^{15+}$ and demonstrate that they lie in the optical region. The transitions are 
insensitive to external perturbations. At the same time they are sensitive to the variation of the
fine structure constant $\alpha$. The fractional accuracy of the
frequency of the transitions can be smaller than
$10^{-19}$, which may provide a basis for atomic clocks of superb accuracy.  
Sensitivity to the variation of $\alpha$ approaches
$10^{-20}$ per year.  
\end{abstract}

\date{ \today }

\maketitle


Building  accurate atomic clocks is important for both technical
applications and fundamental physics. Cesium primary frequency standard which
is currently used to define the SI units of time and length  has fractional accuracy  of
the order $10^{-16}$~\cite{NIST}. Frequency standards based on optical
transitions in neutral atoms trapped in optical lattice aim at fractional
accuracy of $10^{-18}$~\cite{Katori}. Further progress is possible with
clocks using nuclear optical transition~\cite{Th}, or clocks using optical
transitions in highly charged ions~\cite{HCI1,hole,DDF12}. 

One of the important applications of  precise atomic clocks in
fundamental physics is the search for time variation of the fine structure
constant. The possibility for fundamental constants to change in space or time
is suggested by theories unifying gravity with other
interactions~(see, e.g.\cite{Uzan, Flambaum07}).  
The analysis of quasar absorption spectra shows
that there might be a direction in space along which the fine structure
constant $\alpha$ ($\alpha = e^2/\hbar c$) is changing at constant rate over
cosmological distances ($\alpha$-dipole)~\cite{dipole}. Since the
Earth moves in the 
framework of the $\alpha$-dipole, the space variation of $\alpha$ should
manifest itself in terrestrial studies as changing of $\alpha$ in time at
approximate rate of $\dot\alpha/\alpha = 10^{-18}{\rm
  yr}^{-1}$~\cite{BF12}. The best current limit on time-variation of $\alpha$
obtained from comparing Al$^+$ and Hg$^+$ clocks is significantly larger,
$\dot\alpha/\alpha = (-1.6\pm2.3)\times10^{-17}{\rm yr}^{-1}$~\cite{AlHg}.
At least an order of magnitude improvement is needed to verify the
$\alpha$-dipole hypothesis in terrestrial studies. Two or more clocks with
different dependence of clock frequencies on the fine structure constant and
operating on the same level of accuracy ($10^{-18}$ or better) should be
monitored over significant time interval (from few months to few years) to
reveal any variation of $\alpha$ in time. If $\alpha$ is changing, than the
readings of the clocks would shift in time against each other.

It was suggested in Ref.~\cite{HCI1} to use highly charged ions (HCI) for this
purpose. HCIs are less sensitive to external
perturbations than 
neutral atoms (or singly-charged ions) due to their more compact size. Notice that many of HCIs
have optical transitions where standard laser techniques can be used and where
accuracy of the frequency measurements can be possibly even better than in
optical standards using neutral or single-ionized atoms. A number of atomic
transitions in HCI were studied in earlier
works~\cite{HCI1,hole,ions,Cf,crossing} focusing mostly on 
their sensitivity to the variation of the fine structure constant. A detailed
study of the effect of external perturbations on clock frequencies of some
promising HCI systems was recently reported in Ref.~\cite{DDF12}. There it
was shown that 
the relative accuracy on the level of $10^{-19}$ might be possible for HCIs
with the $4f^{12}$ configurations of valence electrons. These systems
are not sensitive to the variation of $\alpha$. However, they can be used as a
reference point against which the frequencies of the transitions sensitive to
the variation of $\alpha$ are monitored.

In present work we study other promising systems which are immune to external perturbations while being sensitive
to the variation of the fine structure constant. This makes them excellent
candidates for  terrestrial studies of the time variation of $\alpha$.
We focus on the $5s$ - $4f$ transitions in  Nd$^{13+}$ and
Sm$^{15+}$ ions. Frequencies of these transitions are in the optical region
due to the $5s$ - $4f$ level crossing which happens for the
isoelectronic sequence of silver atom near $Z=61$ (promethium
ion)~\cite{crossing}.  The first excited states in these ions
($4f_{5/2}$ in Nd$^{13+}$  and the $5s$ in
Sm$^{15+}$) are metastable states because they can only decay to the ground
state via the E3 transition. 
The states are not sensitive to external perturbations. On the
other hand, the $s-f$ 
transitions are the transitions which are most sensitive to the
variation of the fine structure constant~\cite{DFW}. Due to the
inversion of the states from  Nd$^{13+}$ to  Sm$^{15+}$, corresponding
frequencies would move in time in opposite direction if $\alpha$
changes. This further enhances the sensitivity. If the frequencies of
the $5s - 4f_{5/2}$ transition in Nd$^{13+}$ and $4f_{5/2} - 5s$
transition in Sm$^{15+}$ are monitored simultaneously for extended period
of time the sensitivity of the measurements can be on the level
$\delta \alpha/\alpha \sim 10^{-19}$ per year or better.

Note that similar transition in Pm$^{14+}$ is even more sensitive to
variation of $\alpha$. However, in this work we consider only isotopes
with zero nuclear spin to suppress the second-order Zeeman shift. The
promethium atom has no stable isotopes and no long-living isotopes with zero
nuclear spin. 
Therefore we limit ourself in this work to the
Nd$^{13+}$ and Sm$^{15+}$ ions.


The sensitivity of the atomic transitions to the variation of the fine
structure constant can be revealed by varying the value of $\alpha$ in
computer codes. 

We use the correlation potential method~\cite{CPM} to perform the calculations.
The Nd$^{13+}$ and Sm$^{15+}$ ions have a single valence electron
above the Pd-like closed-shell core. Therefore, it is convenient to use the
$V^{N-1}$ approximation in which relativistic Hartree-Fock calculations are
first done for the closed-shell core and states of valence electron are
calculated in the self-consistent field of the frozen core. Correlations
between valence and core electrons are included by constructing the
second-order correlations potential $\hat \Sigma$ and solving the
Hartree-Fock-like equations with an extra operator $\hat \Sigma$ for the states
of external electron
\begin{equation}
(\hat H_0 +\hat \Sigma -\varepsilon_v)\psi_v=0.
\label{eq:Br}
\end{equation}
Here $\hat H_0$ is the relativistic Hartree-Fock Hamiltonian and index
$v$ labels valence orbitals. Solving (\ref{eq:Br}) gives energies and
wave functions for different states of the valence electron which
include correlations (the so called {\em Brueckner orbitals}). 
Many-body perturbation theory and B-spline basis states~\cite{B-spline} are
used to calculate $\hat \Sigma$.

The dependence of the atomic frequencies on the fine structure
constant $\alpha$ the frequencies may be presented as
\begin{equation}
  \omega(x) = \omega_0 + qx, 
\label{eq:w}
\end{equation}
where $x = (\alpha/\alpha_0)^2 -1 $, $\alpha$ is the current value of
the fine structure constant, and $\alpha_0$ is the value of the fine
structure constant at some fixed moment of time, say the beginning of
the observations. Note that we use atomic units which means that the
unit of energy is fixed and does not vary with
$\alpha$. Therefore, equation (\ref{eq:w}) gives comprehensive
description of the frequency dependence on $\alpha$. In the end, only
change of the dimensionless values (e.g., ratio of two frequencies) can be 
studied. Therefore, the actual choice of units cannot affect the
results.
 
The sensitivity coefficient $q$ in (\ref{eq:w}) is calculated by running
the computer code with different values of $\alpha$ and taking numerical
derivative,
\begin{equation}
q=\frac{\omega(0.01)-\omega(-0.01)}{0.02}.
\label{eq:q}
\end{equation}
It follows from (\ref{eq:w}) that the relative change in frequency is related
to the relative change in $\alpha$ by
\begin{equation}
\frac{\delta\omega}{\omega} = \frac{2q}{\omega}\frac{\delta\alpha}{\alpha},
\label{eq:K}
\end{equation}
where $K=2q/\omega$ is the enhancement factor.


\begin{table}
\caption{The $5s$ and $4f$ energy levels and sensitivity coefficients $q$ for
  Nd$^{13+}$ and Sm$^{15+}$ (cm$^{-1}$). $K =2q/\omega$ is the
  enhancement factor.}
\label{t:energy}
\begin{ruledtabular}
\begin{tabular}{llrrr}
\multicolumn{1}{c}{Ion} &\multicolumn{1}{c}{State}
&\multicolumn{1}{c}{Energy} & \multicolumn{1}{c}{$q$} &
\multicolumn{1}{c}{$K$} \\   
\hline
Nd$^{13+}$ & $5s_{1/2}$ &      0 &     0 & 0 \\
           & $4f_{5/2}$ &  58897 & 106000 & 3.4 \\
           & $4f_{7/2}$ &  63613 & 110200 & 3.5 \\
Sm$^{15+}$ & $4f_{5/2}$ &      0 &      0 &  0 \\
           & $4f_{7/2}$ &   6806 &   6300 & 1.9 \\
           & $5s_{1/2}$ &  55675 &-136000 & -4.9 \\
\end{tabular}
\end{ruledtabular}
\end{table}

We chose the Nd$^{13+}$ and  Sm$^{13+}$ ions because the $5s$ -
$4f$ transition for these ions lie in the optical (UV) region. 
Table \ref{t:energy} shows calculated energy levels of the $4f$ and $5s$
states of the ions together with the sensitivity coefficients $q$ (see
(\ref{eq:w}) and (\ref{eq:q})) and the enhancement factor $K$. All other states of
these ions lie very high in the spectrum, far outside of the optical
region. The $5s$ - $4f_{5/2}$ transition is very narrow for both ions.
The lowest-order transitions between the $5s$ and $4f_{5/2}$ states
are the M2 and E3 transitions. The M2 transition is very weak
since it vanishes in the non-relativistic limit. The width of the
lines is determined by the $4f_{5/2}$ - $5s$ E3 transition for
Nd$^{13+}$ and by the $5s$ - $4f_{5/2}$ and $5s$ - $4f_{7/2}$ E3
transitions for Sm$^{15+}$ ion.
The E3 decay rate is given
by (we use atomic units: $\hbar=1$, $m_e=1$, $|e|=1$)
\begin{equation}
\Gamma_b = 0.001693\alpha^7\omega_{ab}^7\frac{\langle a||E3||
  b\rangle^2}{2J_b + 1}. \label{eq:pe3}
\end{equation}
Here $a$ is the ground state and $b$ is the excited metastable state, $\alpha
= 1/137.36$ is the fine structure constant, $\omega_{ab}$ is the frequency of
the transition. Estimates show that for Nd$^{13+}$
$\Gamma_{4f_{5/2}} \approx 1.1 \times 10^{-23}$ a.u. $= 7.4 \times
10^{-8}$ Hz, $1/Q=\Gamma_{4f_{5/2}}/\omega \approx 4.8 \times
10^{-23}$ ($Q$ is the quolity factor), and for
Sm$^{15+}$ \ $\Gamma_{5s} \approx 4.3 \times 10^{-23}$ a.u. $= 2.8 \times
10^{-7}$ Hz, $1/Q=\Gamma_{5s}/\omega \approx 1.7 \times 10^{-22}$ (see
also Table~\ref{t:pol}).

Note, that the $4f$ - $5s$ level crossing~\cite{crossing}  happens at
about $Z=61$ (promethium ion) leading to inversion of the levels order in
samarium ion. As a results, the $q$-coefficients and enhancement factors have
different sign for Nd$^{13+}$ and Sm$^{15+}$. It may be beneficial to
measure the frequency of the $4f_{5/2} - 5s$ transition in Sm$^{15+}$
($\lambda = 170$ nm) against the frequency of the $5s- 4f_{5/2}$
transition in Nd$^{13+}$ ($\lambda = 180$ nm). This would bring
 extra enhancement to the sensitivity of the clock
frequencies to the variation of the fine structure constant. Using
data from Table~\ref{t:energy} one can get
\begin{eqnarray}
  q_1 - q_2 = 242000 \ {\rm cm}^{-1}, \nonumber \\
  \frac{\delta(\omega_1/\omega_2)}{(\omega_1/\omega_2)} = 
  8.3\left(\frac{\delta\alpha}{\alpha}\right).
\label{eq:dw}
\end{eqnarray}
Here $\omega_1$ is the clock frequency for Nd$^{13+}$ and $\omega_2$ is
the clock frequency for Sm$^{15+}$. Eq. (\ref{eq:dw}) shows that if
$\alpha$ drifts, the ratio of the clock frequencies changes
about eight times faster. This means that if the ratio is monitored
for a year with the accuracy of $10^{-19}$, the sensitivity of the
measurements to the variation of $\alpha$ would be close to $10^{-20}$
per year.

\begin{table}
\caption{Parameters of the clock transitions for Nd$^{13+}$ and
  Sm$^{15+}$. Wavelength of the transition ($\lambda$), radiative width of the
  excited clock state ($\Gamma$), static dipole polarizabilities
  ($\alpha(0)$), magnetic dipole polarizabilities ($\gamma_c$),
and magnetic dipole hyperfine structure
  constant ($A, \ g_I \equiv \mu/I$).
  Numbers in square brackets represent powers of 10.} 
\label{t:pol}
\begin{ruledtabular}
\begin{tabular}{l cccc}
 & \multicolumn{2}{c}{Nd$^{13+}$} & \multicolumn{2}{c}{Sm$^{15+}$} \\
 & Ground & Excited & Ground & Excited \\
 & state  & state & state  & state \\
 & $5s$   & $4f_{5/2}$ & $4f_{5/2}$ & $5s$ \\
\hline
$\lambda$ [nm] & \multicolumn{2}{c}{170} & \multicolumn{2}{c}{180} \\
$\Gamma$ [Hz]  & 0 & 7.4[-8] & 0 & 2.8[-7] \\
$\alpha(0)$ [$a_0^3$] & 1.1048 & 0.3701 & 0.2766 & 0.8366 \\
$\gamma_c$ [a.u.] & 0 & 0.144 & 1.344 & 0 \\
$A/g_I$ [MHz] & 123000 & 1070 & 1400 & 154000 \\
\end{tabular}
\end{ruledtabular}
\end{table}

Below we consider various systematic effects which affect the clock transition frequency. 
As discussed in \cite{DDF12} the clock HCI is assumed to be trapped and sympathetically cooled.
\paragraph{Black body radiation shift --}
The frequencies of the clock transitions might be affected by the
black body radiation (BBR) shift. 
The BBR frequency shift at an ambient temperature $T$ can be expressed
as
\begin{equation}
  \frac{\delta\omega}{\omega} = -\left(\frac{T}{T_0}\right)^4
\frac{\Delta\alpha}{2\omega}\left(831.9
  \frac{\rm V}{\rm m}\right)^2,
\label{eq:bbr}
\end{equation}
where $T_0 = 300 \ K$, $\Delta\alpha$ is the difference in the values
of the static dipole polarizabilities of the  clock states.
The polarizability $\alpha_v(0)$ for the state $v$ is given by
\begin{equation}
  \alpha_v(0) = \frac{2}{3(2j_v+1)}\sum_n\frac{\langle v||\mathbf{D}|| n
  \rangle^2}{\varepsilon_n -\varepsilon_v},
\label{eq:pol}
\end{equation}
where $\mathbf{D} =\sum_i e\mathbf{r}_i$ is the electric dipole operator and
summation goes over complete set of states. The results of calculations, which
include Brueckner-type correlations and core polarization effects, are
presented in Table~\ref{t:pol}.
Using (\ref{eq:bbr}) and the data from the table
one can get at room temperature $\delta\omega/\omega = -3.6 \times
10^{-18}$  for Nd$^{13+}$ and $\delta\omega/\omega = 2.9 \times
10^{-18}$ for Sm$^{15+}$. For cryogenic Paul trap operating at the
temperature of liquid helium ($\sim 4 K$) the fractional BBR shift is 
 $\delta\omega/\omega \sim 10^{-25}$  for both ions.

\paragraph{Zeeman shift --}
Clock frequencies are affected by magnetic fields. The first-order Zeeman
shift can be eliminated by averaging the measurements over two virtual clock transitions with opposite g-factors.
Uncontrollable second-oder AC Zeeman shift arises due to misbalances of currents in ion traps. It can be evaluated as
\begin{equation}
\delta E_c = - \frac{1}{2j_v+1}\sum_i \frac{\langle c||\mu||i\rangle^2}{E_i -
  E_c}B^2 \equiv \gamma_c B^2,
\label{eq:Z2}
\end{equation}
The second-order Zeeman shift is negligibly small for the $5s$ states since
there are only strongly forbidden $M1$ transitions in (\ref{eq:Z2}). The shift
for the $4f_{5/2}$ states is dominated by the $4f_{7/2}$ - $4f_{5/2}$
transition within the fine structure doublet. The calculated values of
$\gamma_c$ for these states are presented in Table~\ref{t:pol}.
Using the value of AC magnetic field $B = 5 \times 10^{-8} T$ measured in
Al$^+$/Be$^+$ trap~\cite{AlHg} leads to $\delta\omega/\omega \sim 10^{-26}$.

\paragraph{Electric quadrupole shift--}
Clock frequencies can be affected by coupling of atomic
quadrupole moments to the gradients of trapping electric field. Obviously, this does not
affect the $5s$ state due it's vanishing quadrupole moment, but can shift the $4f_{5/2}$ state. It was suggested
in \cite{DDF12} to use hyperfine structure (hfs) to suppress the quadrupole
shift. However, using isotopes with non-zero nuclear spin leads to
enhanced second-order Zeeman shift. This is because small hfs
intervals enter energy denominators in (\ref{eq:Z2}) increasing the
shift by several orders of magnitude compared to isotopes with zero
nuclear spin. For this reason in this work we propose a different method to
suppress electric quadrupole shift. It uses a linear combination of transition frequencies between states with different projection of
total angular momentum $J$.

The quadrupole shift for a state with total angular momentum $J$ and
its projection $J_z=M$ is given by
\begin{equation}
  \delta E_{JM} ~\sim \frac{3M^2-J(J+1)}{2J(2J-1)}Q\frac{\partial
    E_z}{\partial z} \equiv
  c_{JM}Q\frac{\partial E_z}{\partial z},
\label{eq:at}
\end{equation}
where $Q$ is the electric quadrupole moment.
The dependence of the shift on the projection $M$ is in the prefactor only.
The quadrupole moment $Q$ of the atomic state is defined as
twice the expectation value of the electric quadrupole operator (E2)
in the stretched state 
\begin{equation}
Q=2\langle nJM=J|E2|nJM=J\rangle .
\label{eq:q}
\end{equation}
 
Taking two transitions to states
with different values of $M$ and writing the frequencies as $\omega_M
= \omega_0 + c_{JM}Q (\partial E_z/\partial z)$ one can get
\begin{equation}
  \omega_0 = \frac{\omega_M -
    \omega_{M^{\prime}}c_{JM}/c_{JM^{\prime}}}{ 1 -
    c_{JM}/c_{JM^{\prime}}}.
\label{eq:w0}
\end{equation}
Here $\omega_0$ is the frequency of the transition at zero electric
and magnetic fields. The expression (\ref{eq:w0}) does not depend on
the quadrupole moment nor the gradient of electric field. For better accuracy
it is important to have $c_{JM}$ and $c_{JM^{\prime}}$ as different as
possible. Using $c_{5/2,1/2}=-0.4$ and $c_{5/2,5/2}=0.5$ leads to
\begin{equation}
\omega_0 = \frac{\omega_{1/2} + 0.8\omega_{5/2}}{1.8}.
\label{eq:w01}
\end{equation}.

\paragraph{Other perturbations --}
The performance of  ion clocks can be affected by many other
systematic effects, such as density of the background gases, Doppler
(motion-induced) effects, gravity, etc. Consideration similar to what
was done in our previous work\cite{DDF12} show that none of the
corresponding fractional frequency shift is below the value of $10^{-19}$.

To summarize, the Nd$^{13+}$ and Sm$^{15+}$ HCIs may offer an intriguing possibility for developing clockwork of unprecedented accuracy that is highly sensitive to variation of the fine-structure constant. The $10^{-19}$ fractional accuracy matches projected accuracy of the $^{229}$Th nuclear clock~\cite{Th}, but without complications of radioactivity.


The work was supported in part by the Australian Research Council and U.S. National Science Foundation.

\end{document}